# Spatial- and Frequency-Wideband Effects in Millimeter-Wave Massive MIMO Systems

Bolei Wang, Feifei Gao, Shi Jin, Hai Lin, and Geoffrey Ye Li

*Abstract*—When there are a large number of antennas in massive MIMO systems, the transmitted wideband signal will be sensitive to the physical propagation delay of electromagnetic waves across the large array aperture, which is called the spatial-wideband effect. In this scenario, transceiver design is different from most of the existing works, which presume that the bandwidth of the transmitted signals is not that wide, ignore the spatial-wideband effect, and only address the frequency selectivity. In this paper, we investigate spatial- and frequency-wideband effects, called dual-wideband effects, in massive MIMO systems from array signal processing point of view. Taking mmWave-band communications as an example, we describe the transmission process to address the dual-wideband effects. By exploiting the channel sparsity in the angle domain and the delay domain, we develop the efficient uplink and downlink channel estimation strategies that require much less amount of training overhead and cause no pilot contamination. Thanks to the array signal processing techniques, the proposed channel estimation is suitable for both TDD and FDD massive MIMO systems. Numerical examples demonstrate that the proposed transmission design for massive MIMO systems can effectively deal with the dual-wideband effects.

*Index Terms*—Massive MIMO, mmWave, array signal processing, wideband, spatial-wideband, beam squint, angle reciprocity, delay reciprocity.

## I. INTRODUCTION

Massive MIMO, also known as large-scale MIMO, is formed by equipping hundreds or thousands of antennas at the base station (BS) and can simultaneously serve many users in the same time-frequency band. The large number of antennas in massive MIMO can improve spectrum and energy efficiency, spatial resolution, and network coverage [1]–[3]. In the mean time, millimeter-wave (mmWave) communication makes massive MIMO more attractive since the small antenna size facilitates packing a large number of antennas in a small area at the BS and massive MIMO can combat high path-loss and fading of mmWave channels [3]–[5].

In the past few years, tremendous efforts have been devoted into the development of massive MIMO as well as its application in mmWave-band communications. For example, it has been demonstrated in [6]–[8] that large MIMO systems can

B. Wang and F. Gao are with Tsinghua National Laboratory for Information Science and Technology (TNList), Tsinghua University, Beijing, P. R. China (e-mail: boleiwang@ieee.org, feifeigao@ieee.org).

S. Jin is with the National Communications Research Laboratory, Southeast University, Nanjing 210096, P. R. China (email: jinshi@seu.edu.cn).

H. Lin is with the Department of Electrical and Information Systems, Graduate School of Engineering, Osaka Prefecture University, Sakai, Osaka, Japan (e-mail: hai.lin@ieee.org).

Geoffrey Ye Li is with the School of Electrical and Computer Engineering, Georgia Institute of Technology, Atlanta, GA, USA (email: liye@ece.gatech.edu)

enhance spectral efficiency by several orders of magnitude and simple zero-forcing transceivers are asymptotically optimal. Since channel state information (CSI) is necessary to reach the huge performance gains for massive MIMO, many different channel estimation algorithms [9]–[11] have been designed for uplink channels while downlink channels can be obtained by channel reciprocity for time division duplex (TDD) networks. The calibration error, which may affect the channel reciprocity in the downlink/uplink RF chains, has been also handled in [12]. Downlink channel estimation for frequency division duplex (FDD) networks has been investigated in [13]–[15]. With a huge number of antennas and the corresponding CSI, beamforming can reach a high spatial resolution [16], [17]. To reduce the cost of hardware implementation, hybrid analog-digital beamforming techniques have been developed for mmWave-band massive MIMO systems [18]–[22].

Most of the existing works on massive MIMO are based on the channel model directly extended from the conventional MIMO channel model [5]–[22]. However, as indicated in [23]–[25], there exists a non-negligible time delay across the array aperture for the same data symbol in massive MIMO configuration. Such phenomenon incurs an inherent property of large-size array, called *spatial-wideband (spatial-selective)* effect. In fact, the spatial-wideband effect has been found in radar systems and has already been well studied in the area of array signal processing [26], [27]. In the existing massive MIMO studies, the frequency-selective effect induced by the multipath delay spread has been considered but the spatial-selective effect has been ignored. Therefore, to optimize the performance, the *dual-wideband effects* from both the spatial and the frequency domains should be taken into consideration when designing massive MIMO systems. If the dual-wideband effects are considered, many fundamental issues for massive MIMO, which have been obtained by using the extended conventional MIMO channel models, will need to be reformulated and the previous results will be revised.

In this paper, we investigate the massive MIMO communications by considering the dual-wideband effects from the array signal processing viewpoint. By exploiting the channel sparsity in both angle and delay domains, a massive MIMO channel with the dual-wideband effects is modeled as a function of limited parameters, e.g., the complex gain, the direction of arrival (DOA) or the direction of departure (DOD), and the time delay of each channel path. We then develop channel estimation techniques for both the uplink and downlink cases in massive MIMO systems, which require significantly less amount of training overhead and have no pilot contamination. Furthermore, we will reveal that the angle and the delay domains, or the spatial and the frequency domains, exhibit strong duality in various aspects, which can help the training design as well as the subsequent data transmission. Moreover, the proposed channel estimation strategy is suitable for both TDD and FDD massive MIMO systems due to the angle and the time delay reciprocities. Computer simulation results demonstrate that the proposed channel estimation strategies for massive MIMO systems can effectively deal with the dual-wideband effects while the existing designs cannot.

The rest of this paper is organized as follows. Section II introduces the array signal processing theory as well as the rationale of the spatial-wideband effect. Section III formulates the new channel model of massive MIMO systems considering the dual-wideband effects. Section IV addresses the massive MIMO channel estimation with the dual-wideband effects. Numerical results are provided in Section V, and Section VI concludes this paper.

**Notations:**

| | |
|---|---|
| $(\cdot)^T$ | transpose of a matrix or a vector |
| $(\cdot)^H$ | conjugate transpose or Hermitian operation of a matrix or a vector |
| $(\cdot)^*$ | conjugate of a matrix or a vector |
| $(\cdot)^\dagger$ | pseudo-inversion of a matrix or a vector |
| $\mathbf{I}$ | identity matrix |
| $\mathbf{1}$ | all-ones matrix |
| $\mathbf{0}$ | all-zeros matrix |
| $\circ$ | Hadamard product of two matrices |
| $\otimes$ | Kronecker product of two matrices |
| $\mathbb{E}\{\cdot\}$ | expectation operation |
| $tr(\cdot)$ | matrix trace operation |
| $\|\mathbf{a}\|_2$ | Euclidean norm of vector $\mathbf{a}$ |
| $\mathrm{diag}\{\mathbf{a}\}$ | the diagonal matrix comprising vector $\mathbf{a}$'s elements |
| $\|\mathbf{A}\|_F$ | Frobenius norm of matrix $\mathbf{A}$ |
| $[\mathbf{A}]_{m,n}$ | the $(m, n)$th element of the matrix $\mathbf{A}$ |
| $[\mathbf{A}]_{:,n}$ | the $n$th column of the matrix $\mathbf{A}$ |
| $[\mathbf{A}]_{m,:}$ | the $m$th row of the matrix $\mathbf{A}$ |
| $\mathrm{diag}\{\mathbf{A}\}$ | the vector extracted from diagonal entries of matrix $\mathbf{A}$ |
| $|\mathcal{A}|$ | cardinality of set $\mathcal{A}$ |
| $\mathcal{I}(M)$ | set $\mathcal{I}(M) \triangleq \{0, 1, \dots, M-1\}$ |
| $\Re\{\cdot\}$ | real part of a complex number, vector or matrix |

## II. Spatial-Wideband Effect of Large-Scale Array

In this section, we present the principle of the spatial-wideband effect in large-scale arrays and discuss why it can be ignored in small-scale MIMO systems.

As in Fig. 1, consider a signal from far-field source impinging onto a uniform linear array (ULA) [1] of $M$ antennas with a single direction $\vartheta$. Let the antenna spacing be $d$, the carrier wavelength be $\lambda_c$, and the carrier frequency be $f_c$. Denote $\psi \triangleq \frac{d \sin \vartheta}{\lambda_c}$ for notation simplicity. Then, the time delay between adjacent antennas can be explicitly computed as $\frac{d \sin \vartheta}{c} = \frac{\psi}{f_c}$, where $c$ is the speed of light.

---

[1] In this paper, we employ the ULA to clearly illustrate the nature of the spatial-wideband effect and the corresponding channel characteristics. Nevertheless, the spatial-wideband effect exists in all large-scale antennas regardless of array topology as long as the propagation delay across the array aperture is non-ignorable compared with the symbol duration.

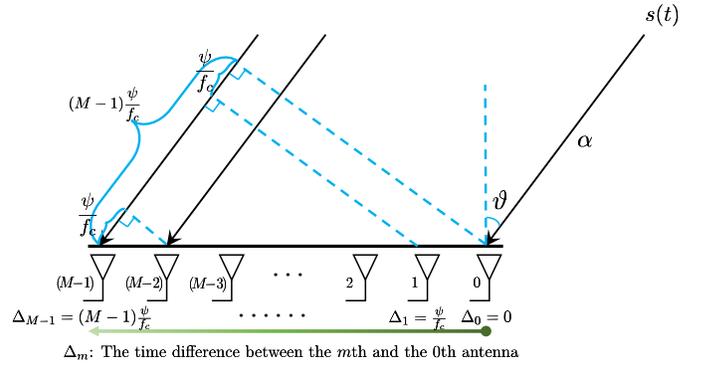

Fig. 1. An illustration of the spatial-wideband effect of a ULA with $M$ antennas.

For a communication system with single-carrier modulation of the symbol period, $T_s$, the baseband signal can be represented as

$$s(t) = \sum_{i=-\infty}^{+\infty} s[i] g(t - iT_s), \tag{1}$$

where $g(t)$ is the pulse shaping function and $s[i]$ is the transmit symbols. The corresponding passband signal can be represented as $\Re\{s(t)e^{j2\pi f_c t}\}$ and the bandwidth of the modulated signal is around $\frac{1}{T_s}$.

Without loss of generality, assume that the first antenna is perfectly synchronized with the source and receives $\Re\{\alpha s(t)e^{j2\pi f_c t}\}$, where $\alpha$ is the complex channel gain. Then, the other antennas will receive a delayed version of $\Re\{\alpha s(t)e^{j2\pi f_c t}\}$.

From Fig. 1, the time interval between the 0th and the $m$th antenna is $m\frac{\psi}{f_c}$, and the passband signal received by the $m$th antenna can be expressed as

$$\Re\left\{\alpha s\left(t - m\frac{\psi}{f_c}\right)e^{j2\pi f_c(t - m\frac{\psi}{f_c})}\right\}$$
$$= \Re\left\{\alpha s\left(t - m\frac{\psi}{f_c}\right)e^{-j2\pi m\psi}e^{j2\pi f_c t}\right\}. \tag{2}$$

Therefore, the equivalent baseband signal received by the $m$th antenna is

$$y_m(t) \triangleq \alpha s\left(t - m\frac{\psi}{f_c}\right)e^{-j2\pi m\psi}. \tag{3}$$

### A. Small Array or Spatial-Narrowband Effect

If the number of antenna $M$ is small, or the transmission bandwidth of $s(t)$ is narrow, then $m\frac{\psi}{f_c} \ll T_s$ and $s(t - m\frac{\psi}{f_c}) \approx s(t)$ hold for $\forall m \in \mathcal{I}(M)$. In this case, the received signal vector from all $M$ antennas can be represented by

$$\mathbf{y}(t) = [y_0(t), y_1(t), \dots, y_{M-1}(t)]^T$$
$$= \alpha s(t)[1, e^{-j2\pi\psi}, \dots, e^{-j2\pi(M-1)\psi}]^T \triangleq \alpha s(t)\mathbf{a}(\psi), \tag{4}$$

where $\mathbf{a}(\psi)$ is known as the *spatial steering vector* pointing towards the angle $\vartheta$. The equivalent baseband channel vector will be $\mathbf{h} = \alpha \mathbf{a}(\psi)$. As a result, the baseband channel can be represented as [11], [29]

$$\mathbf{h} = \sum_{l=0}^{L-1} \alpha_l \mathbf{a}(\psi_l) \delta(t - \tau_l), \tag{5}$$

where $L$ is the number of channel paths,[2] $\psi_l \triangleq \frac{d \sin \vartheta_l}{\lambda_c}$, $\vartheta_l$ denotes the DOA of the $l$th path, $\tau_l$ is the corresponding path delay, and $\delta(\cdot)$ is the Dirac impulse function.

It should be emphasized that the assumption $m \frac{\psi}{f_c} \ll T_s$, which requires the symbol duration $T_s$ to be large enough or the baseband signal $s(t)$ to be relatively narrowband, is a necessary condition for making the small-scale MIMO channel model (5) and other derived models hold.

### B. Large Array and Spatial-Wideband Effect

For wideband massive MIMO system, $M$ is usually very large, and the transmit baseband signal $s(t)$ has wide bandwidth. In this case, the approximation, $s(t - m \frac{\psi}{f_c}) \approx s(t)$, does not hold for the larger indices $m$ any more, making (4) and (5) invalid.

For example, for a ULA system equipped with $M = 128$ antennas and the antenna spacing $d = \lambda_c/2$, when the incident path comes from $\vartheta = 60°$, the propagation delay across the array aperture is $0.58 T_s$ in a typical LTE system with the transmission bandwidth $f_s = 20$ MHz at $f_c = 1.9$ GHz. As for the typical mmWave-band system with $f_s = 1$ GHz at $f_c = 60$ GHz, such delay will be $0.92 T_s$.

More specifically, if $m \frac{\psi}{f_c}$ is comparable to $T_s$ or even larger than $T_s$, then the $m$th antenna would "see" a different transmit symbol from the 0th antenna. This phenomenon is called the *spatial-wideband effect* in array signal processing, which has been well studied in radar signal processing but is seldom discussed in wireless communications where the number of antennas is not large enough or the signal bandwidth is not that wide.

In massive MIMO systems, especially over mmWave-band, the narrowband assumption does not hold any more. In this case, the conventional massive MIMO channel model (5) and other models that directly extended from the conventional models, will be inapplicable. As will be seen in the later numerical results, the algorithms without considering the spatial-wideband effect will suffer from the performance loss when either the number of BS antennas or the transmission bandwidth becomes large. It has been discovered in [30] that the performance of the phased-array hybrid structure will degrade due to the spatial-wideband effect as the bandwidth increases.

---

[2] When there are rich scatters around the BS, especially in low frequency band, many paths would arrive at an approximately identical time delay and formulate one channel tap. In this case, each summand in (5) would be coupled by an integration over the incoming angular spread corresponding to each specific channel tap.

## III. Massive MIMO Channels with Dual-Wideband Effects

As discussed previously, the spatial-wideband effect, as an inherent property of large-scale arrays, must be considered in a real massive MIMO systems. In this section, we will first model the massive MIMO channels with the dual-wideband effects and then discuss its impact on the orthogonal frequency division multiplexing (OFDM) transmission scheme. Moreover, several fundamental characteristics of the dual-wideband channel are also investigated to aid the channel estimation and the user scheduling designs in the next section.

### A. Channel Modeling with Dual-Wideband Effects

Let us consider a massive MIMO system consisting of one BS with an $M$-antenna ULA and $P$ randomly distributed users, each with a single antenna. We will address the dual-wideband effects in OFDM scheme since it is used in most broadband wireless communications protocols. The number of OFDM subcarriers is denoted as $N$. For ease of illustration, we consider the mmWave-band transmission and suppose that there are $L_p$ individual physical paths between the BS and the $p$th user.[3]

Denote $\tau_{p,l,m}$ as the time delay of the $l$th path from the $p$th user to the $m$th antenna, which is not necessarily an integer multiple of $T_s$. Then the received baseband signal at the $m$th antenna from the $p$th user can be expressed as

$$y_{p,m}(t) = \sum_{l=0}^{L_p-1} \bar{\alpha}_{p,l} x_p(t - \tau_{p,l,m}) e^{-j2\pi f_c \tau_{p,l,m}}, \tag{6}$$

where $\bar{\alpha}_{p,l}$ is the corresponding complex channel gain, and $x_p(t)$ is the transmitted signal of the $p$th user. Denote the DOA of the $l$th path of user $p$ as $\vartheta_{p,l}$, and define $\psi_{p,l} \triangleq \frac{d \sin \vartheta_{p,l}}{\lambda_c}$. From Fig. 1, we have

$$\tau_{p,l,m} = \tau_{p,l,0} + m \frac{\psi_{p,l}}{f_c}, \quad m \in \mathcal{I}(M). \tag{7}$$

For notation simplicity, we denote $\tau_{p,l} \triangleq \tau_{p,l,0}$ in the rest of this paper.

With (7), equation (6) can be rewritten as

$$\begin{aligned} y_{p,m}(t) &= \sum_{l=0}^{L_p-1} \bar{\alpha}_{p,l} e^{-j2\pi f_c \tau_{p,l}} x_p\left(t - \tau_{p,l} - m \frac{\psi_{p,l}}{f_c}\right) e^{-j2\pi f_c m \frac{\psi_{p,l}}{f_c}} \\ &= \sum_{l=0}^{L_p-1} \alpha_{p,l} x_p\left(t - \tau_{p,l} - m \frac{\psi_{p,l}}{f_c}\right) e^{-j2\pi m \psi_{p,l}}, \end{aligned} \tag{8}$$

where $\alpha_{p,l} \triangleq \bar{\alpha}_{p,l} e^{-j2\pi f_c \tau_{p,l}}$ is the equivalent channel gain.

---

[3] The extension to low frequency-band communications can be straightwardly made by counting the angular spread caused by local scattering [15], [31].

Then, the uplink *spatial-time* channel of the $p$th user at the $m$th antenna can be modeled as

$$[\mathbf{h}_{ST,p}(t)]_m = \sum_{l=0}^{L_p-1} \alpha_{p,l}[\mathbf{a}(\psi_{p,l})]_m \delta(t - \tau_{p,l,m})$$
$$= \sum_{l=0}^{L_p-1} \alpha_{p,l}[\mathbf{a}(\psi_{p,l})]_m \delta\left(t - \tau_{p,l} - m\frac{\psi_{p,l}}{f_c}\right). \quad (9)$$

Taking the continuous time Fourier transform (CTFT) of (9), we obtain the uplink *spatial-frequency* response of the $p$th user at the $m$th antenna as

$$[\mathbf{h}_{SF,p}(f)]_m = \int_{-\infty}^{+\infty} [\mathbf{h}_{ST,p}(t)]_m e^{-j2\pi f t} dt$$
$$= \sum_{l=0}^{L_p-1} \alpha_{p,l}[\mathbf{a}(\psi_{p,l})]_m e^{-j2\pi f \tau_{p,l,m}}$$
$$= \sum_{l=0}^{L_p-1} \alpha_{p,l} e^{-j2\pi m\psi_{p,l}} e^{-j2\pi f \tau_{p,l}} e^{-j2\pi f m\frac{\psi_{p,l}}{f_c}}. \quad (10)$$

Denote the subcarrier spacing of OFDM as $\eta = \frac{f_s}{N}$ Hz. Then, the spatial-frequency channel coefficients of the $n$th subcarrier can be expressed as $[\mathbf{h}_{SF,p}(n\eta)]_m, \forall n \in \mathcal{I}(N)$. The overall spatial-frequency channel matrix from all $M$ antennas can then be formulated as

$$\mathbf{H}_p = [\mathbf{h}_{SF,p}(0), \mathbf{h}_{SF,p}(\eta), \ldots, \mathbf{h}_{SF,p}((N-1)\eta)]$$
$$= \sum_{l=0}^{L_p-1} \alpha_{p,l} \left(\mathbf{a}(\psi_{p,l})\mathbf{b}^T(\tau_{p,l})\right) \circ \mathbf{\Theta}(\psi_{p,l}), \quad (11)$$

where "$\circ$" denotes the Hadamard product and

$$\mathbf{b}(\tau_{p,l}) \triangleq [1, e^{-j2\pi\eta\tau_{p,l}}, \ldots, e^{-j2\pi(N-1)\eta\tau_{p,l}}]^T \in \mathbb{C}^{N\times 1} \quad (12)$$

can be viewed as *"frequency-domain steering vector"* pointing towards the $l$th delay of the $p$th user. Moreover, $\mathbf{\Theta}(\psi_{p,l})$ is an $M \times N$ matrix whose $(m,n)$th element is

$$[\mathbf{\Theta}(\psi_{p,l})]_{m,n} \triangleq \exp\left(-j2\pi mn\frac{\psi_{p,l}}{f_c}\right),$$
$$m \in \mathcal{I}(M), \quad n \in \mathcal{I}(N). \quad (13)$$

Clearly, the equations (11)-(13) provide a more accurate channel model for large-scale antennas by considering both the spatial- and frequency- wideband effects (dual-wideband effects). In this case, the spatial steering vector $\mathbf{a}(\psi_{p,l})$ is coupled with the frequency-domain steering vector $\mathbf{b}(\tau_{p,l})$, as well as the phase shift matrix $\mathbf{\Theta}(\psi_{p,l})$. We then call (11) the *spatial-frequency wideband (SFW)* channel (*dual-wideband* channel). It should be noted that the *beam squint* effect, which induces the frequency-dependent radiation pattern of a large-scale array [32], is fully considered in the proposed model (11) and expressed by the phase shift matrix $\mathbf{\Theta}(\psi_{p,l})$.

Compared with the conventional small-scale MIMO channel model (5), the phase shift matrix $\mathbf{\Theta}(\psi_{p,l})$ for each path arises due to the spatial-wideband effect. The maximum phase shift inside this matrix can be approximately $\frac{f_s}{f_c}\frac{d}{\lambda_c}M$. In small-scale antennas with small $M$ or in large-scale antennas but with very narrow bandwidth $f_s$, the maximum phase shift is $\frac{f_s}{f_c}\frac{d}{\lambda_c}M \ll 1$ and close to zero such that the phase shift matrix $\mathbf{\Theta}(\psi_{p,l})$ for each path is an approximate all-ones matrix in conformity to the analysis in Sec. II.

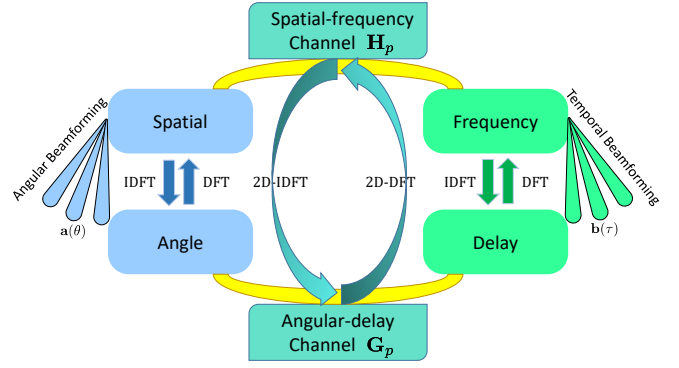

Fig. 2. Duality between the angle domain and the delay domain, as well as between their Fourier transform: spatial domain and frequency domain.

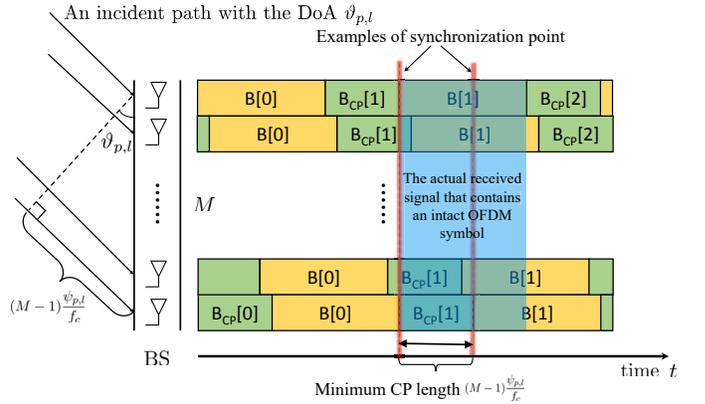

Fig. 3. An illustration of the effect of a single path in SFW channels, where $B[i]$ denotes the $i$th OFDM symbol.

Interestingly, "angle" domain and "delay" domain are coupled in a dual manner in (11). Similar duality also appears in their Fourier transforms: "spatial" domain and "frequency" domain, as illustrated in Fig. 2. For example, the angular beamforming and the temporal beamforming can be realized by weighting the antennas in spatial domain and the subcarriers in the frequency domain, in a similar way. Moreover, the energy of each path in angular-delay domain is diffused for the same level in the angle and the delay dimensions, as will be seen in Theorem 1 in Sec. III-C.

### B. Essential Requirement of CP Length for OFDM

The SFW channel model (11) implicitly requires that the cyclic prefix (CP) of OFDM modulation is sufficiently long to overcome the channel delay at each individual antenna.

To better illustrate how to choose CP length, we first demonstrate the $l$th path of user $p$ in Fig. 3, where $B[i]$ denotes the $i$th time-domain OFDM block of the $p$th user, and $B_{CP}[i]$ represents the corresponding CP. Take $B[1]$ for example. If $\vartheta_{p,l} > 0$, then the CP length to combat *spatial-wideband* effect should be the duration that the $l$th path travel across the whole array, namely $(M-1)\frac{\psi_{p,l}}{f_c}$, as marked by the interval between the two vertical lines in Fig. 3. Considering that the baseband sampling rate is $f_s$, the CP length should be $(M-1)\psi_{p,l}\frac{f_s}{f_c}$. Similarly, if $\vartheta_{p,l} < 0$, then the required CP length is $(M-1)(-\psi_{p,l})\frac{f_s}{f_c}$. Hence, a unified expression of the CP length to combat the spatial-wideband effect can be expressed as $(M-1)|\psi_{p,l}|\frac{f_s}{f_c}$.

Moreover, the time delay of the $l$th path from the user to the array should be combatted by the CP too, which requires additional CP length of $\frac{1}{T_s}\min_{m\in\{0,M-1\}}\tau_{p,l,m}$.[4] Hence, the overall CP length for the $l$th path of the $p$th user should be $(M-1)|\psi_{p,l}|\frac{f_s}{f_c} + \frac{1}{T_s}\min_{m\in\mathcal{I}(M)}\tau_{p,l,m}$.

For the multiuser and multipath scenario, the CP length should be adequate for all incident paths of all users and can then be expressed as

$$\left\lceil \max_{p\in\mathcal{I}(P)}\max_{l\in\mathcal{I}(L_p)}\left[(M-1)|\psi_{p,l}|\frac{f_s}{f_c} + \frac{1}{T_s}\cdot\min_{m\in\mathcal{I}(M)}\tau_{p,l,m}\right]\right\rceil. \tag{14}$$

Nevertheless, a valid CP length should not be a function of the instantaneous DOA. For the overall angle range $\vartheta_{p,l} \in (-\pi/2, \pi/2)$, we have

$$\begin{aligned}
(M-1)|\psi_{p,l}|\frac{f_s}{f_c} &= (M-1)\frac{f_s}{f_c}\frac{d}{\lambda_c}|\sin\vartheta_{p,l}| \\
&\leq (M-1)\frac{f_s}{f_c}\frac{d}{\lambda_c} \xrightarrow{d=\lambda_c/2} \frac{M-1}{2}\frac{f_s}{f_c}.
\end{aligned} \tag{15}$$

Moreover, since $\min_{m\in\mathcal{I}(M)}\tau_{p,l,m} \leq \tau_{p,l,0} = \tau_{p,l}$, a valid CP length could be expressed as

$$N_{CP} = \left\lceil \frac{M-1}{2}\frac{f_s}{f_c} + \frac{1}{T_s}\cdot\max_{p\in\mathcal{I}(P)}\max_{l\in\mathcal{I}(L_p)}\tau_{p,l}\right\rceil. \tag{16}$$

Interestingly, the second item of (16) is exactly the multipath propagation delay seen by the first antenna, which describes the maximum multipath delay spread. Without the spatial-wideband effect, the second item is enough to combat the frequency selectivity. When considering the spatial-wideband effect, the extra CP length, the first term in (16), is additionally required to overcome the time delays across the large array aperture, which is proportional to the transmission bandwidth, $f_s$, and the number of the antennas, $M$, while is inversely proportional to the carrier frequency $f_c$. The spatial-wideband effect will always emerge provided that the transmission bandwidth is sufficiently large when $M \geq 2$.

For the typical parameters in mmWave-band communications, $f_s = 1$ GHz at $f_c = 60$ GHz, the CP length exclusively induced by the spatial-wideband effect is 2 and 9 for the ULAs with the number of antennas $M = 128$ and $M = 1024$, respectively. Hence, the extra burden of large-scale antennas on the length of CP does not seem to be a problematic issue for the existing OFDM protocols. Nevertheless, the subsequent numerical results will demonstrate that the existing algorithms that did not consider the spatial-wideband channel would still suffer from severe performance loss even with sufficient CP length.

### C. Asymptotic Characteristics of SFW Channels

Before characterizing the property of the SFW channel $\mathbf{H}_p$, we first present the asymptotic property of $\boldsymbol{\Theta}(\psi)$, which is proved in Appendix A.

*Lemma 1:* When $M \to \infty, N \to \infty$, if the conditions $\frac{f_s}{f_c}\frac{d}{\lambda_c} < 1$ and $\frac{M-1}{2N}\frac{f_s}{f_c} < 1$ hold,[5] then the two-dimensional inverse discrete Fourier transform (2D-IDFT) of $\boldsymbol{\Theta}(\psi)$ is a block-sparse matrix with all nonzero elements staying inside a square region, i.e.,

$$\lim_{M,N\to\infty}\left[\mathbf{F}_M^H\boldsymbol{\Theta}(\psi)\mathbf{F}_N^*\right]_{i,k} = \begin{cases} \text{nonzeros} & (i,k)\in\mathcal{A}_1 \\ 0 & (i,k)\notin\mathcal{A}_1 \end{cases} \tag{17}$$

where $\mathbf{F}_M$ is the normalized $M$-dimensional DFT matrix with its $(p,q)$th entry $[\mathbf{F}_M]_{p,q} \triangleq e^{-j\frac{2\pi}{M}pq}/\sqrt{M}$, and

$$\mathcal{A}_1 \triangleq \begin{cases} \left\{(i,k)\in\mathbb{Z}\,\big|\,0\leq i < \frac{f_s}{f_c}\frac{d\sin\vartheta}{\lambda_c}M,\right. \\ \qquad \left. 0\leq k < \frac{f_s}{f_c}\frac{d\sin\vartheta}{\lambda_c}M\right\}, \vartheta\in\left[0,\frac{\pi}{2}\right] \\ \left\{(i,k)\in\mathbb{Z}\,\big|\,M - \frac{f_s}{f_c}\frac{d\sin\vartheta}{\lambda_c}M \leq i < M,\right. \\ \qquad \left. N - \frac{f_s}{f_c}\frac{d\sin\vartheta}{\lambda_c}M \leq k < N\right\}, \vartheta\in\left[-\frac{\pi}{2},0\right) \end{cases}.$$

Based on Lemma 1, we can prove the following theorem in Appendix B.

*Theorem 1:* When $M \to \infty, N \to \infty$, the *angular-delay channel* $\mathbf{G}_p \triangleq \mathbf{F}_M^H\mathbf{H}_p\mathbf{F}_N^*$ is a sparse matrix and only possesses $L_p$ non-zero square regions, each corresponding to one of channel paths.

Compared with the conventional MIMO-OFDM channel, the proposed dual-wideband channel in angular-delay domain performs totally different. Without the spatial-wideband effect, the phase shift matrix $\boldsymbol{\Theta}(\psi_{p,l})$ does not exist or is an all-ones matrix. Correspondingly, each path manifests an impulse in angular-delay domain. With the spatial-wideband effect, Lemma 1 and Theorem 1 indicate that $\boldsymbol{\Theta}(\psi_{p,l})$ diffuses the energy of each path from an impulse to a square region. Different paths manifest different sizes of square dependent upon the path DOA. An example of 6-path SFW channel is shown in Fig. 4, where the location of each square reflects the DOA and time delay of each path based on Theorem 1.

---

[4]Here, $\frac{1}{T_s}\min_{m\in\{0,M-1\}}\tau_{p,l,m}$ means the earliest arriving time of the $l$th path onto one of the $M$ antennas.

[5]The first inequality naturally holds, since the signal bandwidth is far less than its carrier frequency in wireless communication systems and the antenna spacing is usually smaller than half of the carrier wavelength in massive MIMO. The second inequality is definitely true under the proposed CP design (16).

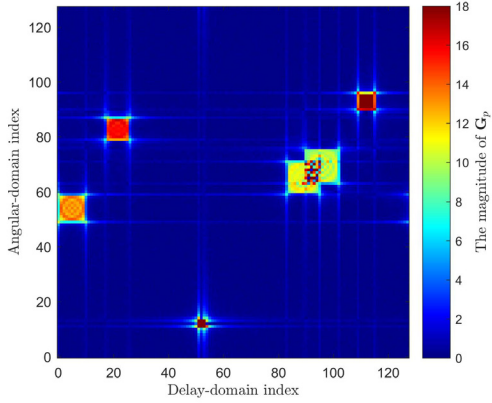

(a) Grayscale of angular-delay domain

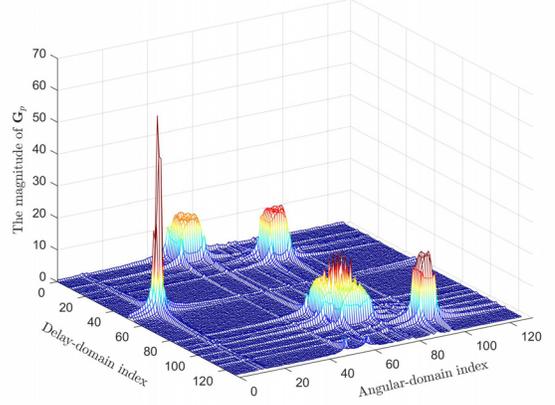

(b) Surface plot of angular-delay domain

Fig. 4. An illustration of a 6-path SFW channel, where $M = 128$, $N = 128$, $d/\lambda_c = 0.5$, $f_s/f_c = 0.2$, and SNR $= 10$dB.

### D. Sparse Channel Representation and Angular-Delay Orthogonality

Define the vectorizing SFW channel $\mathbf{h}_p \triangleq \text{vec}(\mathbf{H}_p)$ as

$$
\begin{aligned}
\mathbf{h}_p &= \sum_{l=0}^{L_p-1} \alpha_{p,l} \left[ \text{vec}\left(\mathbf{\Theta}(\psi_{p,l})\right) \circ \text{vec}\left(\mathbf{a}(\psi_{p,l})\mathbf{b}^T(\tau_{p,l})\right) \right] \\
&= \sum_{l=0}^{L_p-1} \alpha_{p,l} \text{diag}(\text{vec}(\mathbf{\Theta}(\psi_{p,l}))) \left(\mathbf{b}(\tau_{p,l}) \otimes \mathbf{a}(\psi_{p,l})\right) \\
&\triangleq \sum_{l=0}^{L_p-1} \alpha_{p,l} \mathbf{p}(\psi_{p,l}, \tau_{p,l}) \in \mathbb{C}^{MN \times 1},
\end{aligned}
\tag{18}
$$

where $\mathbf{p}(\psi_{p,l}, \tau_{p,l}) \triangleq \text{diag}\left(\text{vec}\left(\mathbf{\Theta}(\psi_{p,l})\right)\right)\left(\mathbf{b}(\tau_{p,l}) \otimes \mathbf{a}(\psi_{p,l})\right)$ is the corresponding item and serves as the basis vector for spanning $\mathbf{h}_p$. The following theorem proved in Appendix C indicates the orthogonality of the basis vectors of $\mathbf{h}_p$.

*Theorem 2:* When $M \to \infty, N \to \infty$, the following property holds

$$
\lim_{M,N \to \infty} \frac{1}{MN} \mathbf{p}(\psi_1, \tau_1)^H \mathbf{p}(\psi_2, \tau_2) = \begin{cases} 1 & \psi_1 = \psi_2, \tau_1 = \tau_2 \\ 0 & \text{otherwise} \end{cases}.
\tag{19}
$$

Based on Theorem 2, if any two users do not share the same path (paths with both the same DOA and the same time delay), then their vectorized SFW channels are asymptotically orthogonal. This phenomenon is called the *angular-delay orthogonality*, which indicates the orthogonality for users at different locations, or for users at the same location but with different path delays. Interestingly, as shown in Fig. 4, two squares in Fig. 4(a) corresponding to two different paths may partially overlap with each other, which, however, does not break the angular-delay orthogonality in terms of Theorem 2 unless the two squares locate at the exact identical position.

An important application of Theorem 2 is in channel estimation and user scheduling. By exploiting the angular-delay orthogonality, orthogonal users in angular-delay domain can be simultaneously scheduled without pilot contamination or mutual interference, as demonstrated in the next section.

## IV. Channel Estimation for Dual-Wideband mmWave Massive MIMO Systems

The above discussion has demonstrated that the SFW channel model of the massive MIMO system is not just an extension of the traditional MIMO channel model. This critical difference demands for redesign of most communication strategies, such as channel estimation, signal detection, beamforming, precoding, user scheduling, etc. Due to the space limitation, we present a simple channel estimation algorithm in this section to deal with the dual-wideband effects by the aid of array signal processing.

### A. Preamble for Initial Uplink Channel Estimation

In the preamble phase, we apply the conventional least square MIMO-OFDM channel estimation algorithm [39] for each antenna at the BS. Denote $\widehat{\mathbf{H}}_p$ as the uplink preamble channel between the $p$th user and the BS. The main purpose of the preamble is to obtain the initial DOA and the time delay of each path for each user and facilitate the subsequent uplink and downlink channel estimations with a small number of pilot resources.

### B. Extracting Angular-Delay Signature

From Theorem 1, $\vartheta_{p,l}$, $\tau_{p,l}$, and $L_p$ can be immediately obtained from the non-zero square of $\widehat{\mathbf{G}}_p = \mathbf{F}_M^H \widehat{\mathbf{H}}_p \mathbf{F}_N^*$. However, when $M$ and $N$ are finite in practice, the region of the non-zero square will be expanded due to the power leakage effect [15], [31]. Hence, $\vartheta_{p,l}$, $\tau_{p,l}$, and $L_p$ should be obtained by a more sophisticatedly designed way.

Denote

$$
\mathbf{\Psi}_M(\Delta\psi_{p,l}) = \text{diag}\left(1, e^{j\Delta\psi_{p,l}}, \ldots, e^{j(M-1)\Delta\psi_{p,l}}\right)
\tag{20}
$$

---

**Step 1:** *Detect the number of path for each user.* As shown in Fig. 4, each channel path exhibits a cluster of points in a non-zero square region of $\widehat{\mathbf{G}}_p = \mathbf{F}_M^H \widehat{\mathbf{H}}_p \mathbf{F}_N^*$. Calculate the number of clusters as the estimation of $L_p$.

**Step 2:** *Determine the coarse estimation for each path.* In the cluster of points corresponding to the $l$th path of the $p$th user, locate the two-tuple grid index $(m_{p,l}, n_{p,l})$ of the dominant value. Then, $(\frac{m_{p,l}}{M}, \frac{n_{p,l}}{N\eta})$ can be viewed as the coarse estimation of the angular-delay signature $(\psi_{p,l}, \tau_{p,l})$.

**Step 3:** *Search the angular-delay signature.* For the $l$th path of the $p$th user, search the angular-delay signature near $(\frac{m_{p,l}}{M}, \frac{n_{p,l}}{N\eta})$ by determining

$$(\hat{\psi}_{p,l}, \hat{\tau}_{p,l}) = \operatorname*{argmax}_{(\psi_{p,l}, \tau_{p,l}) \in U(\frac{m_{p,l}}{M}, \frac{n_{p,l}}{N\eta})} \left| \mathbf{f}_{M,\check{\psi}_{p,l}}^H \boldsymbol{\Psi}_M(\Delta\psi_{p,l}) \left[ \widehat{\mathbf{H}}_p \circ \boldsymbol{\Theta}^*(\psi_{p,l}) \right] \boldsymbol{\Psi}_N(\Delta\tau_{p,l}) \mathbf{f}_{N,\check{\tau}_{p,l}}^* \right|$$

where $U(a, b)$ is defined as the neighbourhood region of the point $(a, b)$, and $\check{\hat{\psi}}_{p,l}, \Delta\hat{\psi}_{p,l}, \check{\hat{\tau}}_{p,l}$ and $\Delta\hat{\tau}_{p,l}$ come from the definition (24)-(26). After performing the search for all of $\sum_{p=1}^P L_p$ paths, the angular-delay signatures of all users are obtained as $\mathcal{B}_p = \{ (\hat{\psi}_{p,l}, \hat{\tau}_{p,l}) \mid l \in \mathcal{I}(L_p) \}, \forall p \in \mathcal{I}(P)$.

---

and

$$\boldsymbol{\Psi}_N(\Delta\tau_{p,l}) = \operatorname{diag}\left(1, e^{j\Delta\tau_{p,l}}, \dots, e^{j(N-1)\Delta\tau_{p,l}}\right) \quad (21)$$

as the angular rotation matrix and the delay rotation matrix, respectively, and the *angular-delay rotation* as

$$\mathbf{H}_{p,l}^r = \boldsymbol{\Psi}_M(\Delta\psi_{p,l}) \left(\mathbf{H}_{p,l} \circ \boldsymbol{\Theta}^*(\psi_{p,l})\right) \boldsymbol{\Psi}_N(\Delta\tau_{p,l}). \quad (22)$$

The angular-delay rotation is to rotate $\mathbf{H}_{p,l}$ in both angle domain and delay domain such that all channel power of a path will be concentrated on one entry and the corresponding rotation can be used to estimate $\psi_{p,l}$ and $\tau_{p,l}$, as shown in the following theorem, proved in Appendix D.

*Theorem 3:* The 2D-IDFT of $\mathbf{H}_{p,l}^r$ will have one single non-zero entry

$$\left[\mathbf{G}_{p,l}^r\right]_{\check{\psi}_{p,l},\check{\tau}_{p,l}} = \left[\mathbf{F}_M^H \mathbf{H}_{p,l}^r \mathbf{F}_N^*\right]_{\check{\psi}_{p,l},\check{\tau}_{p,l}} = \sqrt{MN}\alpha_{p,l}. \quad (23)$$

when

$$\Delta\psi_{p,l} = 2\pi\psi_{p,l} - \frac{2\pi}{M}\ddot{\psi}_{p,l} \in (-\pi/M, \pi/M], \quad (24)$$

$$\Delta\tau_{p,l} = 2\pi\eta\tau_{p,l} - \frac{2\pi}{N}\ddot{\tau}_{p,l} \in (-\pi/N, \pi/N], \quad (25)$$

where

$$\ddot{\psi}_{p,l} = \frac{1 + \operatorname{sgn}\psi_{p,l}}{2}\lceil M\psi_{p,l}\rceil + \frac{1 - \operatorname{sgn}\psi_{p,l}}{2}\lceil M\psi_{p,l} + M\rceil \in \mathcal{I}(M), \quad (26)$$

$$\ddot{\tau}_{p,l} = \lceil N\eta\tau_{p,l}\rceil \in \mathcal{I}(N). \quad (27)$$

Let us define the *angular-delay signature* of the $p$th user as

$$\mathcal{B}_p = \left\{ (\psi_{p,l}, \tau_{p,l}) \mid l \in \mathcal{I}(L_p) \right\}, \quad (28)$$

i.e, a user is characterized by its DOAs and time delays. Based on Theorem 3, the concrete steps to estimate the angular-delay signatures are shown in Table I.

### C. Uplink Channel Estimation by Soft Grouping

The channel state information should be re-estimated for every channel coherence time. From [40]–[42], a user's physical position changes much slower than the channel variation. Therefore, it is reasonable that the angular-delay signature of a user remains unchanged within tens of channel coherence times. With the aid of previously obtained angular-delay signatures, only the channel gains need to be re-estimated. Furthermore, the angular-delay sparsity allows users with different angular-delay signatures to train by the same pilot sequence, without causing any pilot contamination any more, as indicated in Theorem 2.

We here develop a *soft grouping* strategy to embrace all users to train simultaneously. Specifically, we purposely adjust each user's training time, such that the angular-delay signatures of all users are different from each other. Denote the angular-delay signature of the $p$th user after time adjustment as $\tilde{\mathcal{B}}_p = (\psi_{p,l}, \tilde{\tau}_{p,l})$. To cope with practically finite $M$ and $N$, we set guard intervals among all $\tilde{\mathcal{B}}_p$, i.e.,

$$\tilde{\mathcal{B}}_p \cap \tilde{\mathcal{B}}_r = \varnothing \quad \text{and} \quad \operatorname{dist}(\tilde{\mathcal{B}}_p, \tilde{\mathcal{B}}_r) \geq \Omega, \quad (29)$$

where $\Omega$ is a pre-determined guard interval, and $\operatorname{dist}(\tilde{\mathcal{B}}_p, \tilde{\mathcal{B}}_r)$ is the distance between any two angular-delay signatures, defined as

$$\operatorname{dist}(\tilde{\mathcal{B}}_p, \tilde{\mathcal{B}}_r) \triangleq \min_{\substack{l_1 \in \mathcal{I}(L_p) \\ l_2 \in \mathcal{I}(L_r)}} \left\| [M\psi_{p,l_1}, N\eta\tilde{\tau}_{p,l_1}]^T - [M\psi_{p,l_2}, N\eta\tilde{\tau}_{p,l_2}]^T \right\|_2. \quad (30)$$

Note that the proposed soft grouping strategy is different from the conventional hard grouping one [11], [15], [31] in that the latter requires users with overlapped signature to transmit over completely non-overlapped time intervals while ours allows users to transmit in an "interlock" way by purposely postpone certain users such that (29) can be satisfied. Hence, the hard grouping strategy can be regarded as a special case of the proposed soft grouping one if we postpone users to stay in a completely non-overlapped manner.

To present a simple channel estimation algorithm for the time being, we assume that all users send pilot "1" over all their carriers of the training block. Then the received signal at all antennas from the entire OFDM block can be expressed as an $M \times N$ matrix

$$\mathbf{Y}_U = \sum_{p \in \mathcal{I}(P)} \sqrt{E_p}\mathbf{H}_p + \mathbf{W}_U, \quad (31)$$

where $\sqrt{E_p}$ is the training power constraint for user $p$, and $\mathbf{W}_U \in \mathbb{C}^{M \times N}$ is the independent additive white Gaussian noise matrix with each element distributed as $\mathcal{CN}(0, \sigma_n^2)$.

By exploiting the angular-delay orthogonality of SFW channels and the obtained signatures of all users, the complex channel gain of each path can be simply updated as[6]

$$\hat{\alpha}_{p,l} \approx \frac{1}{MN\sqrt{E_p}} \mathbf{p}(\psi_{p,l}, \tilde{\tau}_{p,l})^H \text{vec}(\mathbf{Y}_U). \quad (32)$$

The uplink channel of the $p$th user can then be reconstructed by the updated complex gain with the corresponding angular-delay signatures as

$$\hat{\mathbf{H}}_p^U = \sum_{l \in \mathcal{I}(L_p)} \hat{\alpha}_{p,l} \left( \mathbf{a}(\psi_{p,l}) \mathbf{b}^T(\tilde{\tau}_{p,l}) \right) \circ \mathbf{\Theta}(\psi_{p,l}), \quad (33)$$

where $\psi_{p,l}$ and $\tilde{\tau}_{p,l}$ are obtained during the preamble and the soft-grouping process.

From (11) or (18), the high-dimensional SFW channel can be represented in a sparse form determined by the $3L_p$ parameters, $\vartheta_{p,l}$ (or the equivalent $\psi_{p,l}$), $\tau_{p,l}$, and $\alpha_{p,l}$.

### D. Downlink Channel Representation

It has been shown in [43], [44] that the physical DOAs $\vartheta_{p,l}$ and path delays $\tau_{p,l}$ are roughly the same for the uplink and downlink transmission in FDD systems[7], which is called the *angular-delay reciprocity*. Denote the downlink carrier frequency as $f_c^D$ and the corresponding carrier wavelength as $\lambda_c^D$. According to the angular-delay reciprocity, $\psi_{p,l}$ in the downlink channel can be directly obtained from

$$\psi_{p,l}^D = \frac{d \sin \vartheta_{p,l}}{\lambda_c^D} = \frac{d \sin \vartheta_{p,l}}{\lambda_c} \frac{f_c^D}{f_c} = \frac{f_c^D}{f_c} \psi_{p,l}. \quad (34)$$

Similar to the uplink case, we can use the soft grouping method by adjusting the path delay from $\tau_{p,l}$ to $\tilde{\tau}_{p,l}^D$ and then obtain the downlink signature $\tilde{\mathcal{B}}_p^D$. The adjustment obeys the same rule as (29) expect for replacing $(\psi_{p,l}, \tilde{\tau}_{p,l})$ by $(\psi_{p,l}^D, \tilde{\tau}_{p,l}^D)$.

Denote the downlink channel from the BS towards the $p$th user as $\left( \mathbf{H}_p^D \right)^H \in \mathbb{C}^{N \times M}$. Similar to (11), the downlink spatial-frequency channel can be modeled as

$$\mathbf{H}_p^D = \sum_{l=0}^{L_p - 1} \beta_{p,l} \left( \mathbf{a}(\psi_{p,l}^D) \mathbf{b}^T(\tilde{\tau}_{p,l}^D) \right) \circ \mathbf{\Theta}(\psi_{p,l}^D) \in \mathbb{C}^{M \times N}. \quad (35)$$

As a result, we only need to estimate the corresponding $\beta_{p,l}$ in (35) to obtain the downlink channel.

---

[6]As long as the angular-delay signatures of users change slowly, one can also obtain the updated angular-delay signatures from (31) with Algorithm 1, where the non-zero blocks of all users can be distinguished thanks to the proposed soft grouping method.

[7]This characteristic of electromagnetic wave holds when the interval between the downlink and the uplink frequency is within several gigahertz (GHz).

### E. Downlink Channel Estimation

Let us assume that the users contain the same number of paths $L_M$ and the training is performed over $L_M$ blocks. Denote $\mathbf{h}_p^D \triangleq \text{vec}(\mathbf{H}_p^D) \in \mathbb{C}^{MN \times 1}$. The downlink channel between the BS and the $p$th user can be expressed as

$$(\mathbf{h}_p^D)^H = \sum_{l=0}^{L_p - 1} (\beta_{p,l})^* \mathbf{p}^H(\psi_{p,l}^D, \tilde{\tau}_{p,l}^D) = \beta_p^H \mathbf{P}_p^H, \quad (36)$$

where $\beta_p = [\beta_{p,0}, \beta_{p,1}, \ldots, \beta_{p,L_p-1}]^T$ is the downlink channel gain vector and $\mathbf{P}_p \in \mathbb{C}^{MN \times L_M}$ is defined as

$$\mathbf{P}_p \triangleq \left[ \mathbf{p}(\psi_{p,0}^D, \tilde{\tau}_{p,0}^D), \mathbf{p}(\psi_{p,1}^D, \tilde{\tau}_{p,1}^D), \ldots, \mathbf{p}(\psi_{p,L_p-1}^D, \tilde{\tau}_{p,L_p-1}^D) \right]. \quad (37)$$

Obviously, if we choose the beamforming matrix for the $p$th user as $\mathbf{B}_p = \frac{1}{MN} \mathbf{P}_p$, then the BS will formulate beams separately pointing towards each path of the $p$th user and yield the optimal estimation of $\beta_p$. Hence, the overall beamforming matrix for all users can be expressed as

$$\mathbf{B}^D = \sum_{p \in \mathcal{I}(P)} \mathbf{B}_p. \quad (38)$$

Denote $\mathbf{S}$ as an $L_M \times L_M$ training matrix. On the $n$th subcarrier of the $q$th block, $n \in \mathcal{I}(N), q \in \mathcal{I}(L_M)$, the training symbol sent by the $m$th antenna is $\left[ \mathbf{B}^D[\mathbf{S}]_{:,q} \right]_{nM+m}$. Then the received signal of the $p$th user at all $N$ subcarriers in the $q$th block can be expressed as

$$\mathbf{y}_{p,q} = (\bar{\mathbf{H}}_p^D)^H \mathbf{B}^D [\mathbf{S}]_{:,q} + \mathbf{w}_{p,q} \in \mathbb{C}^{N \times 1}, \quad (39)$$

where $\mathbf{w}_{p,q} \in \mathbb{C}^{N \times 1}$ is the corresponding noise, and

$$\bar{\mathbf{H}}_p^D \triangleq \begin{bmatrix} [\mathbf{H}_p^D]_{:,0} & \mathbf{0} & \cdots & \mathbf{0} \\ \mathbf{0} & [\mathbf{H}_p^D]_{:,1} & \cdots & \mathbf{0} \\ \vdots & \vdots & \ddots & \vdots \\ \mathbf{0} & \mathbf{0} & \mathbf{0} & [\mathbf{H}_p^D]_{:,N-1} \end{bmatrix} \in \mathbb{C}^{MN \times N}. \quad (40)$$

The $p$th user then sums the received signals from all subcarriers and obtains an $L_M \times 1$ vector $\mathbf{y}_p$ over $L_M$ blocks as

$$\mathbf{y}_p^H = \left[ \sum_{n=0}^{N-1} [\mathbf{y}_{p,0}]_n, \sum_{n=0}^{N-1} [\mathbf{y}_{p,1}]_n, \ldots, \sum_{n=0}^{N-1} [\mathbf{y}_{p,L_M-1}]_n \right]. \quad (41)$$

It can be further calculated as

$$\begin{aligned} \mathbf{y}_p^H &= (\mathbf{h}_p^D)^H \mathbf{B}^D \mathbf{S}^H + \bar{\mathbf{w}}_p^H \\ &= \beta_p^H \mathbf{P}_p^H \mathbf{B}_p \mathbf{S}^H + \sum_{r \neq p} \beta_p^H \mathbf{P}_p^H \mathbf{B}_r \mathbf{S}^H + \bar{\mathbf{w}}_p^H, \end{aligned} \quad (42)$$

where $\bar{\mathbf{w}}_p^H$ is the equivalent noise vector.

Since the second item of (42) is asymptotically zero, i.e.,

$$\lim_{M,N \to \infty} \mathbf{P}_p^H \mathbf{B}_r = \delta[p-r] \mathbf{I}_{L_M}, \quad \forall p, r \in \mathcal{I}(P), \quad (43)$$

the downlink channel gain can be estimated as

$$\hat{\beta}_p^H = \frac{1}{L_M} \mathbf{y}_p^H \mathbf{S} \approx \beta_p^H + \frac{1}{L_M} \bar{\mathbf{w}}_p^H \mathbf{S}. \quad (44)$$

Then, the downlink SFW channel can be rebuilt as

$$(\widehat{\mathbf{h}}_p^D)^H = \hat\beta_p^H \mathbf{P}_p^H \approx (\mathbf{h}_p^D)^H + \frac{1}{L_M}\bar{\mathbf{w}}_p^H \mathbf{S}\mathbf{P}_p^H. \quad (45)$$

With the proposed scheme, the number of estimated parameters for each user in the downlink training phase is exactly the number of its physical multipaths, i.e., the utilization of the sparsity is made in the extreme. Users do not need to know the angular-delay signatures of itself and everything is handled at the BS directly, which significantly reduces the feedback cost.

## V. SIMULATION RESULTS

In this section, we present several examples to validate the proposed methods and demonstrate the superiority over the existing methods. The BS is equipped with ULA and the antenna spacing is half of the downlink carrier wavelength. The uplink and the downlink carrier frequencies are set as $f_c = 58$ GHz and $f_c^D = 60$ GHz, respectively. All of $P = 10$ users are equipped with a single antenna and are uniformly distributed throughout the cell. Each user contains $1 \sim 6$ channel paths and thus the proposed approach needs to estimate up to 6 pairs of angle and delay for each user. The guard interval is set as $\Omega = 10$. The absolute mean-square errors (MSEs) of the estimated angle and the path delay are defined as

$$\text{MSE}_\vartheta = \mathbb{E}\big\{|\hat\vartheta_{p,l} - \vartheta_{p,l}|^2\big\}, \quad \text{MSE}_\tau = \mathbb{E}\big\{|\hat\tau_{p,l} - \tau_{p,l}|^2\big\},$$

respectively. The normalized MSEs of the channel gain, the uplink spatial-frequency channel matrix, and the downlink spatial-frequency channel vector are defined as

$$\text{NMSE}_\alpha = \frac{\mathbb{E}\big\{|\hat\alpha_{p,l} - \alpha_{p,l}|^2\big\}}{\mathbb{E}\big\{|\alpha_{p,l}|^2\big\}},$$

$$\text{NMSE}_U = \frac{\mathbb{E}\big\{\|\widehat{\mathbf{H}}_p - \mathbf{H}_p\|_F^2\big\}}{\mathbb{E}\big\{\|\mathbf{H}_p\|_F^2\big\}},$$

$$\text{NMSE}_D = \frac{\mathbb{E}\big\{\|\widehat{\mathbf{h}}_p - \mathbf{h}_p\|_F^2\big\}}{\mathbb{E}\big\{\|\mathbf{h}_p\|_F^2\big\}},$$

respectively.

In the first example, we compare the proposed method with two conventional frequency-wideband-only algorithms: the on-grid CS method [9] and the gridless CS method [38]. Fig. 5 compares the downlink performance of all three methods under various transmission bandwidths. The carrier frequency is under mmWave band and set as $f_c^D = 60$ GHz. The number of the antennas is set as $M = 128$ and the number of subcarriers is set as $N = 128$. From Fig. 5, for the regular mmWave-band parameters, e.g., $f_s = 1$ GHz, both the frequency-wideband-only algorithms fails, as expected. When the signal bandwidth decreases, the on-grid CS method does not improve due to the grid mismatch effect (or the power leakage effect [15]). Nevertheless, the gridless CS method gradually improves its performance and approaches the proposed dual-wideband method at $f_s = 20$ MHz, in which case the spatial-wideband effect is negligible and the channel can be treated as spatially narrowband. Note that, the propagation delays across the array of $f_s = 20$ MHz, $f_s = 100$ MHz, and $f_s = 1$ GHz are $0.021T_s$, $0.106T_s$, and $1.06T_s$, respectively, which implies that even a small time delay across array would deteriorate the conventional methods.

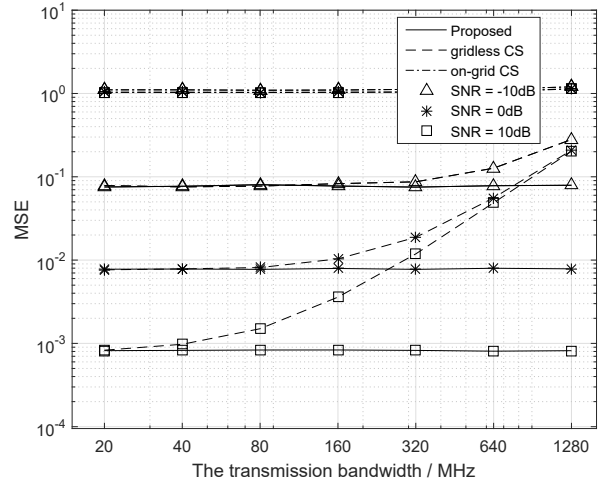

Fig. 5. Channel estimation NMSEs of the proposed method and the conventional methods versus bandwidths, with $M = 128$, $N = 128$, and $f_c^D = 60$ GHz.

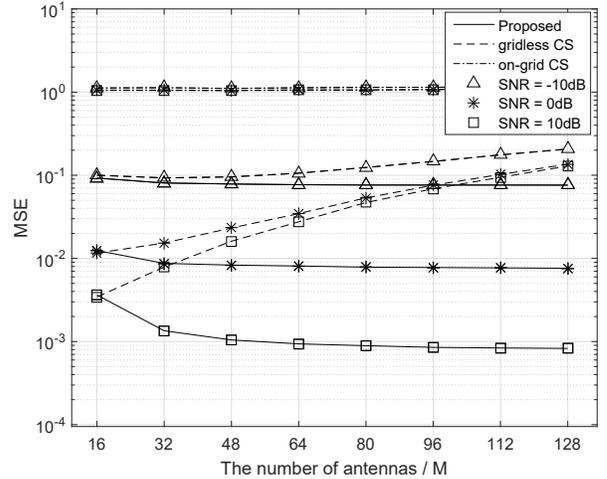

Fig. 6. Channel estimation NMSEs of the proposed method and the conventional methods versus the number of antennas, with $N = 128$, $f_s = 1$ GHz, and $f_c^D = 60$ GHz.

We next show the downlink channel estimation performance of different methods with various number of antennas in Fig. 6. The transmission bandwidth is kept as $f_s = 1$ GHz at $f_c = 60$ GHz. The number of subcarrier is set as $N = 128$. The power is normalized over the number of antennas, $M$, such that the energy received by users keeps constant as $M$ increases. From the figure, when $M$ increases, say above 16, the spatial-wideband effect gradually becomes severe and the performance of gridless CS method degrades. The on-grid method consistently performs poorly due to its serious power leakage effect. Nevertheless, the proposed method works well for all values of $M$. Note that, the propagation delays across the array of $M = 16$, $M = 64$ and $M = 128$ are $0.125T_s$, $0.525T_s$ and $1.06T_s$, respectively.

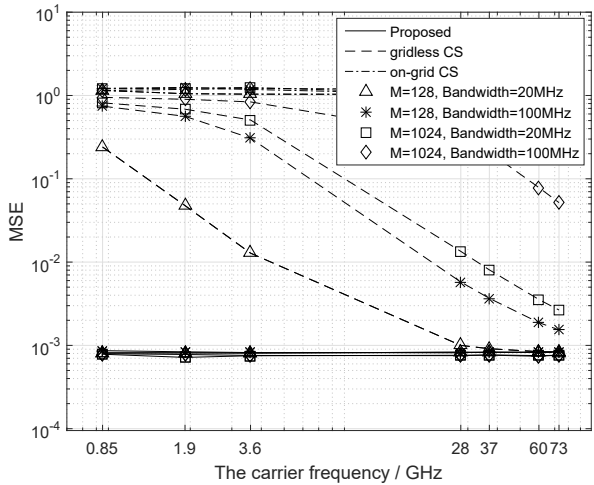

Fig. 7. Channel estimation NMSEs of the proposed method and the conventional methods versus the carrier frequency, with $N = 128$ and SNR $= 10$ dB.

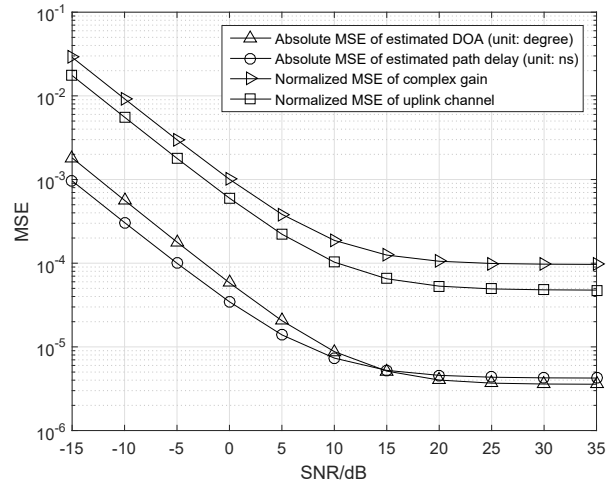

Fig. 8. MSEs of angle and delay and NMSEs of channel in uplink estimation by the proposed strategy, with $M = 128$, $N = 128$, and $f_s = 1$ GHz, respectively.

Fig. 7 further compares the downlink performance of the three methods under various transmission bandwidths and carrier frequencies. The number of subcarriers is set as $N = 128$, and SNR is fixed in 10 dB. The number of the antennas is set as $M = 128$ and $M = 1024$, respectively. The transmission bandwidth is set as $f_s = 20$ MHz and $f_s = 100$ MHz, respectively. From Fig. 7, even for the typical parameters, e.g., $f_s = 20$ MHz at $f_c = 1.9$ GHz, the spatial-wideband effect is clearly observed when $M$ is 128. Hence, for these typical system parameters, the traditional algorithms developed for spatial-narrowband transmission cannot well estimate the channels. Note that the maximum propagation delay across the ULA under this parameter is $0.668T_s$ when $M = 128$, or $5.38T_s$ when $M = 1024$, respectively.

From the above comparison, when the number of antennas in a ULA is fewer than 16, the spatial-wideband effect can be ignored. It indicates that the conventional MIMO channel model is still applicable on a few current massive MIMO prototypes with the $8 \times 8$ and the $16 \times 16$ planar arrays at those above-mentioned system parameters. Nevertheless, for a massive MIMO system with the antenna array containing more than 16 antennas in one of dimension(s), the spatial-wideband effect cannot be neglected and should be carefully treated.

Next, we demonstrate the uplink and the downlink performances of the proposed approach on different conditions.

Fig. 8 shows the MSEs of the proposed uplink channel estimation versus SNR with $M = 128$ and $N = 128$, respectively. The angular-delay signature for each user is first obtained from the preamble phase. With the estimated angular-delay signature, the users are scheduled by the proposed soft grouping method and then the channel gains are updated by one training block in the subsequent transmission. As SNR increases, the estimation accuracy of all parameters increases but will meet error floors at relatively high SNR. Such error floors are caused by the slight angular-delay non-orthogonality since $M$ and $N$ are finite.

Fig. 9 illustrates the MSEs of the proposed uplink and down-link channel estimation with different numbers of antennas, $M$. The number of subcarriers is fixed in $N = 128$. From the figure, more antennas will provide better performance for both uplink and downlink channel estimations. In the low SNR region, different numbers of antennas perform similarly because the overall transmit powers from all cases are kept the same. Nevertheless, more antennas at the BS will present the better beamforming such that the error floor progressively decreases at the high SNR region.

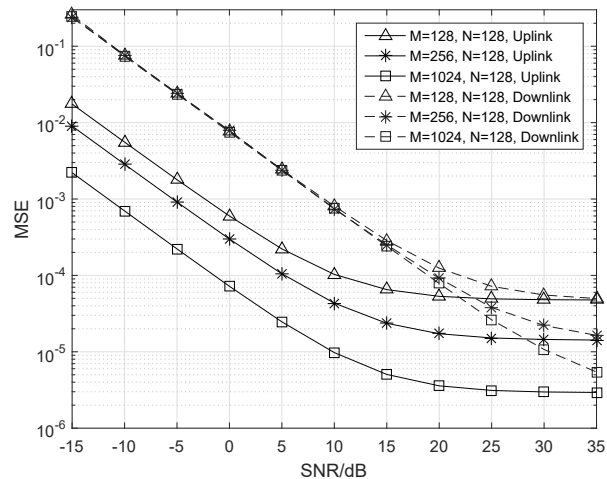

Fig. 9. NMSE of uplink and downlink channel estimations versus SNR, with $f_s = 1$ GHz, $N = 128$, and $M = 128, 256, 1024$, respectively.

Fig. 10 illustrates the MSEs of the proposed uplink and downlink channel estimation with different numbers of subcarriers $N$. The number of BS antennas is kept as $M = 128$. The uplink channel estimation performs similarly as in Fig. 9. For downlink channel estimation, larger $N$ will provide more received energy since we did not normalize the power in one OFDM block, and hence presents better channel estimation. Nevertheless, both uplink and downlink channel estimation will meet the same error floor that is determined by angular-delay non-orthogonality since $M$ and $N$ are finite.

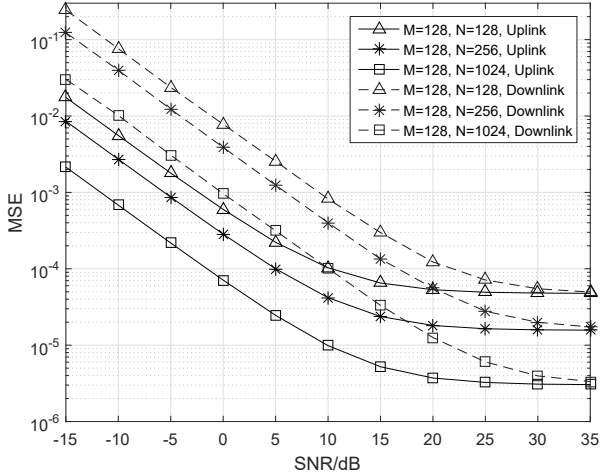

Fig. 10. NMSE of uplink and downlink channel estimations versus SNR, with $f_s = 1$ GHz, $M = 128$, and $N = 128, 256, 1024$, respectively.

## VI. CONCLUSIONS

In this paper, we have re-investigated the massive MIMO communications by taking into account the real propagation of the electromagnetic wave. We have first identified the *spatial-wideband effect* caused by massive number of antennas from array signal processing viewpoint. Then, we have described the mmWave-band channels as functions of limited user-specific parameters, e.g., the complex gain, the DOA/DOD, and the time delay of each channel path. For ease of illustration, we have presented a simple uplink and downlink channel estimation strategy that could successfully obtain the channel information with significantly less amount of training overhead, whereas more sophisticated designs under various optimization have been left as open problems. Moreover, the proposed channel estimation strategy is suitable for both TDD and FDD massive MIMO systems thanks to angular-delay reciprocity. Numerical examples have demonstrated that the conventional massive MIMO algorithms could fail even for regular system parameters while the proposed one is effective under various system parameters if the dual-wideband effects are considered.

## APPENDIX A
## PROOF OF LEMMA 1

Let us first consider the case $\vartheta \in [0, \pi/2]$, and it can be easily computed that

$$
\left[ \mathbf{F}_M^H \mathbf{\Theta}(\psi) \right]_{i,n} = \frac{1}{\sqrt{M}} \sum_{m=0}^{M-1} e^{j\frac{2\pi}{M} im} e^{-j2\pi mn\eta \frac{\psi}{f_c}}
$$

$$
= \frac{1}{\sqrt{M}} \sum_{m=0}^{M-1} e^{-j2\pi m(n\eta \frac{\psi}{f_c} - \frac{i}{M})} = \frac{1}{\sqrt{M}} \sum_{m=0}^{M-1} e^{-j2\pi m(\frac{n}{N} \frac{f_s}{f_c} \psi - \frac{i}{M})}
$$

$$
= \frac{1}{\sqrt{M}} \frac{1 - e^{-j2\pi M(\frac{n}{N} \frac{f_s}{f_c} \psi - \frac{i}{M})}}{1 - e^{-j2\pi(\frac{n}{N} \frac{f_s}{f_c} \psi - \frac{i}{M})}}
$$

$$
= \frac{1}{\sqrt{M}} \frac{e^{j\pi M(\frac{n}{N} \frac{f_s}{f_c} \psi - \frac{i}{M})} - e^{-j\pi M(\frac{n}{N} \frac{f_s}{f_c} \psi - \frac{i}{M})}}{e^{j\pi(\frac{n}{N} \frac{f_s}{f_c} \psi - \frac{i}{M})} - e^{-j\pi(\frac{n}{N} \frac{f_s}{f_c} \psi - \frac{i}{M})}} e^{-j\pi(M-1)(n\eta \frac{\psi}{f_c} - \frac{i}{M})}
$$

$$
= \frac{1}{\sqrt{M}} \frac{\sin\left(\pi M\left(\frac{n}{N} \frac{f_s}{f_c} \psi - \frac{i}{M}\right)\right)}{\sin\left(\pi\left(\frac{n}{N} \frac{f_s}{f_c} \psi - \frac{i}{M}\right)\right)} e^{-j\pi(M-1)(\frac{n}{N} \frac{f_s}{f_c} \psi - \frac{i}{M})}.
$$
(46)

When $M \to \infty$, there is

$$
\lim_{M \to \infty} \left| \left[ \mathbf{F}_M^H \mathbf{\Theta}(\psi) \right]_{i,n} \right| = \sqrt{M} \delta\left( \frac{n}{N} \frac{f_s}{f_c} \psi - \frac{i}{M} \right), \; n \in \mathcal{I}(N).
$$
(47)

When $\frac{f_s}{f_c} \frac{d}{\lambda_c} < 1$, there is

$$
0 \le \frac{n}{N} \frac{f_s}{f_c} \psi = \frac{n}{N} \frac{f_s}{f_c} \frac{d \sin \vartheta}{\lambda_c} \le \frac{f_s}{f_c} \frac{d \sin(\vartheta)}{\lambda_c} \le \frac{f_s}{f_c} \frac{d}{\lambda_c} < 1.
$$
(48)

Hence, (46) is distributed as

$$
\lim_{M \to \infty} \mathbf{F}_M^H \mathbf{\Theta}(\psi) = \left[ \mathbf{U}_{N \times \left(\frac{f_s}{f_c} \frac{d \sin \vartheta}{\lambda_c} M\right)} \; \mathbf{0}_{N \times \left(M - \frac{f_s}{f_c} \frac{d \sin \vartheta}{\lambda_c} M\right)} \right]^T
$$
(49)

where $\mathbf{U}$ is a non-zero matrix. Similarly, by calculating

$$
[\mathbf{\Theta}(\psi) \mathbf{F}_N^*]_{m,k} = \frac{1}{\sqrt{N}} \sum_{n=0}^{N-1} e^{-j2\pi mn\eta \frac{\psi}{f_c}} e^{j\frac{2\pi}{N} nk}
$$

$$
= \frac{1}{\sqrt{N}} \sum_{n=0}^{N-1} e^{-j2\pi n(m\eta \frac{\psi}{f_c} - \frac{k}{N})}
$$

$$
= \frac{1}{\sqrt{N}} \frac{\sin\left(\pi N\left(m\eta \frac{\psi}{f_c} - \frac{k}{N}\right)\right)}{\sin\left(\pi\left(m\eta \frac{\psi}{f_c} - \frac{k}{N}\right)\right)} e^{-j\pi(N-1)(m\eta \frac{\psi}{f_c} - \frac{k}{N})},
$$
(50)

we obtain

$$
\lim_{N \to \infty} \left| [\mathbf{\Theta}(\psi) \mathbf{F}_N^*]_{m,k} \right| = \sqrt{N} \delta\left( m\eta \frac{\psi}{f_c} - \frac{k}{N} \right), \; m \in \mathcal{I}(M).
$$
(51)

According to (14), there is

$$
0 \le m\eta \frac{\psi}{f_c} \le (M-1)\eta \frac{\psi}{f_c} = \frac{M-1}{N} \frac{f_s}{f_c} \frac{d \sin \vartheta}{\lambda_c}
$$

$$
< \frac{1}{N} \left[ \frac{(M-1)}{2} \frac{f_s}{f_c} \right] < 1, \quad (52)
$$

and we have

$$
\lim_{N \to \infty} \mathbf{\Theta}(\psi) \mathbf{F}_N^* = \left[ \mathbf{L}_{M \times \frac{f_s}{f_c} \frac{d \sin \vartheta}{\lambda_c} M} \; \mathbf{0}_{M \times \left(N - \frac{f_s}{f_c} \frac{d \sin \vartheta}{\lambda_c} M\right)} \right],
$$
(53)

where $\mathbf{L}$ is a non-zero matrix.

From (49) we know $\lim_{M \to \infty} \mathbf{F}_M^H \mathbf{\Theta}(\psi) \mathbf{A}$ would have the last $\left(M - \frac{f_s}{f_c} \frac{d \sin \vartheta}{\lambda_c} M\right)$ rows being $\mathbf{0}_{\left(M - \frac{f_s}{f_c} \frac{d \sin \vartheta}{\lambda_c} M\right) \times N}$ for any matrix $\mathbf{A}$, while from (53) we know $\lim_{N \to \infty} \mathbf{B} \mathbf{\Theta}(\psi) \mathbf{F}_N^*$ would have the last $\left(N - \frac{f_s}{f_c} \frac{d \sin \vartheta}{\lambda_c} M\right)$ columns being $\mathbf{0}_{M \times \left(N - \frac{f_s}{f_c} \frac{d \sin \vartheta}{\lambda_c} M\right)}$ for any matrix $\mathbf{B}$. Hence, $\mathbf{F}_M^H \mathbf{\Theta}(\psi) \mathbf{F}_N^*$ must have the the following asymptotic structure:

$$
\lim_{M,N \to \infty} \mathbf{F}_M^H \mathbf{\Theta}(\psi) \mathbf{F}_N^* = \left( \begin{array}{c|c} \mathbf{C}_{\frac{f_s}{f_c} \frac{d \sin \vartheta}{\lambda_c} M \times \frac{f_s}{f_c} \frac{d \sin \vartheta}{\lambda_c} M} & \mathbf{0} \\ \hline \mathbf{0} & \mathbf{0} \end{array} \right),
$$
(54)

where $\mathbf{C}$ is the corresponding non-zero matrix.

Similar result can be derived for $\vartheta \in [-\pi/2, 0)$, and this completes the proof.

## APPENDIX B
## PROOF OF THEOREM 1

Due to the linearity of the 2D-IDFT operation, we only need to investigate the sparse effect of the $l$th path. Define $\mathbf{H}_{p,l} = \alpha_{p,l} \left( \mathbf{a}(\psi_{p,l}) \mathbf{b}^T(\tau_{p,l}) \right) \circ \mathbf{\Theta}(\psi_{p,l})$, whose element is

$$[\mathbf{H}_{p,l}]_{m,n} = \alpha_{p,l} \left[ \mathbf{\Theta}(\psi_{p,l}) \right]_{m,n} e^{-j2\pi \left[ \frac{m}{M}(M\psi_{p,l}) + \frac{n}{N}(N\eta\tau_{p,l}) \right]}. \quad (55)$$

Applying the shift property of 2D-IDFT [33], $\mathbf{G}_{p,l} \triangleq \mathbf{F}_M^H \mathbf{H}_{p,l} \mathbf{F}_N^*$ can be immediately demonstrated satisfying the following asymptotical property

$$\lim_{M,N \to \infty} [\mathbf{G}_{p,l}]_{i,k} = \lim_{M,N \to \infty} [\mathbf{F}_M^H \mathbf{H}_{p,l} \mathbf{F}_N^*]_{i,k}$$
$$= \begin{cases} \text{nonzeros} & (i,k) \in \mathcal{A}_2 \\ 0 & (i,k) \notin \mathcal{A}_2 \end{cases}, \quad (56)$$

where

$$\mathcal{A}_2 = \big\{ (i,k) \in \mathbb{Z} \mid i = \text{mod}(i_1 + M\psi_{p,l}, M), $$
$$k = \text{mod}(k_1 + N\eta\tau_{p,l}, N), \forall (i_1, k_1) \in \mathcal{A}_1 \big\} \quad (57)$$

and $\text{mod}(a, m) \in \mathcal{I}(m)$ is defined as the remainder after division of $a$ by $m$.

## APPENDIX C
## PROOF OF THEOREM 2

First, consider the situation when $\vartheta_1 \neq \vartheta_2$, namely $\psi_1 \neq \psi_2$. Then there is

$$\mathbf{p}(\psi_1, \tau_1)^H \mathbf{p}(\psi_2, \tau_2)$$
$$= \sum_{n=0}^{N-1} \sum_{m=0}^{M-1} e^{j2\pi n\eta\tau_1} e^{j2\pi m\psi_1} e^{j2\pi mn\eta \frac{\psi_1}{f_c}} e^{-j2\pi n\eta\tau_2}$$
$$\times e^{-j2\pi m\psi_2} e^{-j2\pi mn\eta \frac{\psi_2}{f_c}}$$
$$= \sum_{n=0}^{N-1} e^{-j2\pi n\eta(\tau_2 - \tau_1)} \sum_{m=0}^{M-1} e^{-j2\pi m(\psi_2 - \psi_1)} e^{-j2\pi mn\eta(\frac{\psi_2}{f_c} - \frac{\psi_1}{f_c})}$$
$$= \sum_{n=0}^{N-1} e^{-j2\pi n\eta(\tau_2 - \tau_1)} \frac{\sin(\pi M[(\psi_2 - \psi_1) + n\eta(\psi_2 - \psi_1)/f_c])}{\sin(\pi[(\psi_2 - \psi_1) + n\eta(\psi_2 - \psi_1)/f_c])}$$
$$\times e^{-j\pi(M-1)[(\psi_2 - \psi_1) + n\eta(\psi_2 - \psi_1)/f_c]}. \quad (58)$$

Considering the range of $\psi_1$ and $\psi_2$, and using the following equivalence relationships

$$\psi_2 - \psi_1 + n\eta(\psi_2 - \psi_1)/f_c = 0$$
$$\iff \psi_2 \left( 1 + \frac{n}{N} \frac{f_s}{f_c} \right) - \psi_1 \left( 1 + \frac{n}{N} \frac{f_s}{f_c} \right) = 0$$
$$\iff \psi_2 = \psi_1, \quad (59)$$

we can obtain

$$\lim_{M \to \infty} \frac{1}{M} \frac{\sin(\pi M[(\psi_2 - \psi_1) + n\eta(\psi_2 - \psi_1)/f_c])}{\sin(\pi[(\psi_2 - \psi_1) + n\eta(\psi_2 - \psi_1)/f_c])}$$
$$= \delta(\psi_2 - \psi_1), \quad \forall n \in \mathcal{I}(N). \quad (60)$$

For the case $\psi_1 = \psi_2$, (19) can be simplified from (58)–(60) as

$$\lim_{M,N \to \infty} \frac{1}{MN} \mathbf{p}^H(\psi_1, \tau_1) \mathbf{p}(\psi_2, \tau_2)$$
$$= \lim_{N \to \infty} \frac{1}{N} \sum_{n=0}^{N-1} e^{-j2\pi n\eta(\tau_2 - \tau_1)} = \delta(\tau_2 - \tau_1). \quad (61)$$

Combining (60) and (61) completes the proof.

## APPENDIX D
## PROOF OF THEOREM 3

When $\psi_{p,l} \geq 0$, it can be computed that

$$\left[ \mathbf{G}_{p,l}^r \right]_{i,k}$$
$$= \left[ \mathbf{F}_M^H \mathbf{\Psi}_M(\Delta\psi_{p,l}) \left( \mathbf{H}_{p,l} \circ \mathbf{\Theta}^*(\psi_{p,l}) \right) \mathbf{\Psi}_N(\Delta\tau_{p,l}) \mathbf{F}_N^* \right]_{i,k}$$
$$= \frac{\alpha_{p,l}}{\sqrt{MN}} \sum_{n=0}^{N-1} e^{j\frac{2\pi}{N} nk} \times$$
$$\sum_{m=0}^{M-1} e^{j\frac{2\pi}{M} im} e^{jm\Delta\psi_{p,l}} e^{-j2\pi m\psi_{p,l}} e^{-j2\pi n\eta\tau_{p,l}} e^{jn\Delta\tau_{p,l}}$$
$$= \frac{\alpha_{p,l}}{\sqrt{MN}} \sum_{n=0}^{N-1} e^{j\frac{2\pi}{N} nk} e^{-jn(2\pi\eta\tau_{p,l} - \Delta\tau_{p,l})} \times$$
$$\sum_{m=0}^{M-1} e^{j\frac{2\pi}{M} im} e^{-jm(2\pi\psi_{p,l} - \Delta\psi_{p,l})}$$
$$= \frac{\alpha_{p,l}}{\sqrt{MN}} \sum_{n=0}^{N-1} e^{j\frac{2\pi}{N} n(k - \ddot{\tau}_{p,l})} \sum_{m=0}^{M-1} e^{j\frac{2\pi}{M} m(i - \ddot{\psi}_{p,l})}$$
$$= \sqrt{MN} \alpha \cdot \delta(i - \ddot{\psi}_{p,l}) \delta(k - \ddot{\tau}_{p,l}). \quad (62)$$

Similar result can be obtained when $\psi_{p,l} < 0$.